\documentclass[journal=jacsat,manuscript=article]{achemso}

\usepackage{chemformula} 
\usepackage[T1]{fontenc} 
\usepackage[font=small]{caption} 

\author{Andrea Barbiero}
\affiliation{Toshiba Europe Limited, 208 Science Park, Milton Road, Cambridge, CB4 0GZ, UK}
\email{andrea.barbiero@toshiba.eu}
\author{Ginny Shooter}
\affiliation{Toshiba Europe Limited, 208 Science Park, Milton Road, Cambridge, CB4 0GZ, UK}
\author{Tina M\"{u}ller}
\affiliation{Toshiba Europe Limited, 208 Science Park, Milton Road, Cambridge, CB4 0GZ, UK}
\author{Joanna Skiba-Szymanska}
\affiliation{Toshiba Europe Limited, 208 Science Park, Milton Road, Cambridge, CB4 0GZ, UK}
\author{R. Mark Stevenson}
\affiliation{Toshiba Europe Limited, 208 Science Park, Milton Road, Cambridge, CB4 0GZ, UK}
\author{Lucy E. Goff}
\affiliation{Cavendish Laboratory, University of Cambridge, Madingley Road, Cambridge, CB3 0HE, United Kingdom}
\author{David A. Ritchie}
\affiliation{Cavendish Laboratory, University of Cambridge, Madingley Road, Cambridge, CB3 0HE, United Kingdom}
\author{Andrew J. Shields}
\affiliation{Toshiba Europe Limited, 208 Science Park, Milton Road, Cambridge, CB4 0GZ, UK}

\title[An \textsf{achemso} demo]
  {Polarization-selective enhancement of telecom wavelength quantum dot transitions in an elliptical bullseye resonator}

\abbreviations{IR,NMR,UV}

\begin{document}


\begin{abstract}
\noindent 
Semiconductor quantum dots are promising candidates for the generation of nonclassical light.
Coupling a quantum dot to a device capable of providing polarization-selective enhancement of optical transitions is highly beneficial for advanced functionalities such as efficient resonant driving schemes or applications based on optical cyclicity.
Here, we demonstrate broadband polarization-selective enhancement by coupling a quantum dot emitting in the telecom O-band to an elliptical bullseye resonator. 
We report bright single-photon emission with a degree of linear polarization of 96\%, Purcell factor of 3.9, and count rates up to 3 MHz. Furthermore, we present a measurement of two-photon interference without any external polarization filtering and demonstrate compatibility with compact Stirling cryocoolers by operating the device at temperatures up to 40 K. 
These results represent an important step towards practical integration of optimal quantum dot photon sources in deployment-ready setups.
\end{abstract}


\section{Introduction}
\noindent Single-photon sources are an important building block for the advancement of emerging quantum technologies such as quantum key distribution \cite{Gisin.2002,Xu.2020, Couteau.2023}, quantum metrology \cite{Couteau.2023b} and quantum computing \cite{Kok.2007, Slussarenko.2019}. For most applications, an optimal source should efficiently deliver indistinguishable single photons in a well-defined spatial and polarization mode \cite{Thomas.2021b}. 
Among the several platforms investigated over the last decades, semiconductor quantum dots (QDs) \cite{Shields.2007} coupled to Purcell microcavities have shown excellent quantum-optical properties (high purity and indistinguishability) and state-of-the-art performance in terms of brightness and repetition rate \cite{He.2013, Ding.2016, Senellart.2017, Arakawa.2020}.
Moreover, their ability to emit directly in the telecom O-band (1260 nm - 1360 nm) \cite{Zinoni.2006,Ward.2014} or C-band (1530 nm - 1565 nm) \cite{Portalupi.2019, Anderson.2021}  is crucial for the development of long-distance quantum communication networks \cite{Kimble.2008, Xu.2020} thanks to the low-loss windows of standard silica fibers at those wavelengths. 

Multiple semiconductor devices have shown the potential to serve as optimal cavity structures for the generation of single photons at high rates \cite{Somaschi.2016, Gregersen.2016, Liu.2018, Uppu.2020, Tomm.2021}. Among these, circular Bragg gratings (CBGs, also known as bullseye resonators)  \cite{Davanco.2011, Liu.2019, Wang.2019} have recently emerged as promising candidates thanks to a unique combination of moderate Purcell enhancement and high extraction efficiency in a broad spectral range. Moreover, they do not require demanding fine features and guarantee good robustness against common imperfections, such as tilted side walls or spatial displacement of the QD emitter \cite{Rickert.2019, Barbiero.2022, Bremer.2022}. 
Singly-charged QDs in CBGs have already shown impressive count rates and first-lens efficiencies \cite{Kolatschek.2021, Barbiero.2022b, Nawrath.2023}, and combining those performance with a polarization-selective enhancement \cite{Unitt.2005, Lee.2014} would be beneficial for multiple applications.
First, it is worth noting that the most efficient QD source currently available relies on polarization selectivity to separate the flux of single photons from the driving laser and avoid a 50\% loss in the collection efficiency \cite{Tomm.2021}. Furthermore, a polarization-selective device can induce cycling transitions, which are a crucial ingredient for the generation of photonic cluster states with protocols based on time-bin encoded qubits \cite{Lee.2019, Appel.2021}. Finally, applications such as quantum key distribution and any interference based experiments rely on single photons generated into a well-defined polarization state.
Polarized emission from a CBG has been observed when the QD is displaced from the centre of the cavity \cite{Peniakov.2023}. However this approach, which requires accurate deterministic fabrication, may have a negative impact on the Purcell enhancement because the emitter is not positioned in the region of highest confinement.
An alternative strategy relies on the introduction of a small ellipticity to break the symmetry of the cavity. The first report of a QD coupled to an elliptical Bragg grating (EBG) has already displayed impressive performance, with a Purcell factor of $\sim$ 15.7 and a polarized single-photon efficiency of $\sim$ 56\% \cite{Wang.2019c}. Nonetheless, the operating wavelength around 880 nm prevents its integration with the existing optical fiber infrastructure, and an EBG source emitting directly in the telecom regime has not been demonstrated yet.

In this work, we report on the fabrication of EBG cavities operating in the telecom O-band.
By investigating the photoluminescence (PL) signal of a self-assembled InAs/GaAs QD embedded in one of those devices we demonstrate that the cavity strongly enhances one of the two orthogonal polarization components, resulting in bright single-photon emission with a degree of linear polarization > 96\%, Purcell factor of 3.9, and count rates up to 3 MHz.
To demonstrate the effect of strong polarization selective enhancement, we report a measurement of two-photon interference without any external polarization filtering. Finally, we show stable operation of the linearly polarized single-photon source up to a temperature of 40 K for compatibility with state-of-the-art compact cryocoolers.

\section{Results and Discussion}
%
\begin{figure}[h]
\includegraphics[width=1\textwidth]{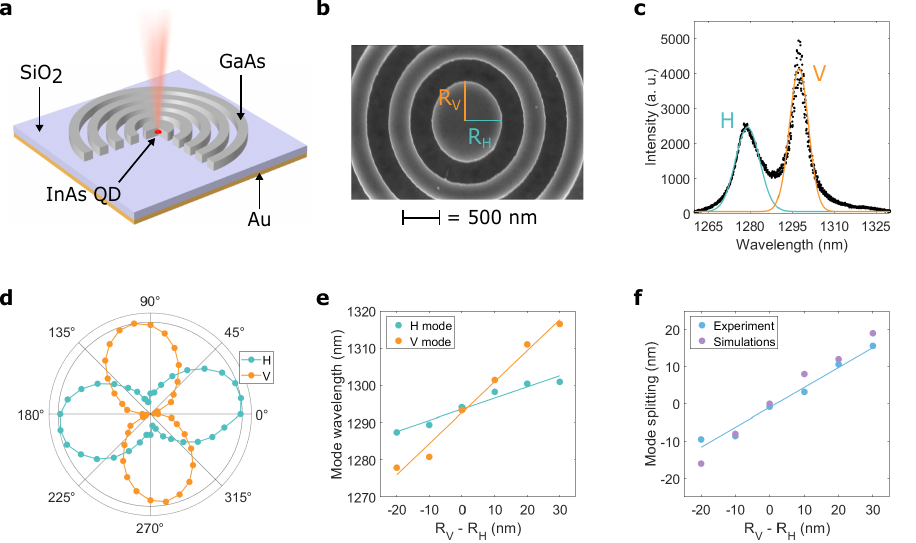}
\caption{\textbf{Structure and characterization of the elliptical Bragg grating cavities.} (a) Illustration of the EBG device with a single InAs QD in the center of the cavity, an insulating SiO\textsubscript{2} layer and a backside Au mirror. 
(b) SEM top view of an exemplary EBG cavity. The labels R\textsubscript{\textit{H}} and R\textsubscript{\textit{V}} indicate the major and minor radius of the elliptical central disk, respectively.
(c) Two non-degenerate cavity modes of an EBG measured under strong above-band CW laser excitation. The solid lines represent the Gaussian fit of the data. The modes are labelled \textit{H} and \textit{V}, with FHWM\textsubscript{\textit{H}} = 9.68 nm, FHWM\textsubscript{\textit{V}} = 6.49 nm, and a splitting of 17.73 nm. 
(d) Polarization-resolved measurements of the two orthogonal cavity modes. The degree of linear polarization of the \textit{H} and \textit{V} modes is 79.4\% and 96.3\%, respectively. 
(e) Central wavelength of the \textit{H} and \textit{V} modes measured on a series of 6 EBG devices with fixed R\textsubscript{\textit{H}} = 515 nm and variable R\textsubscript{\textit{V}}. The solid lines represent the linear fit of the data. 
(f) Measured splitting of the orthogonal cavity modes from (e) and corresponding values predicted by FEM simulations. The solid lines represent the linear fit of the data. 
\label{Fig1} }
\end{figure}
\noindent Similarly to conventional bullseye cavities, an elliptical Bragg grating consists of concentric periodic trenches etched in a semiconductor membrane around a central disk containing the QD emitter (Figure \ref{Fig1}a). This device also includes an insulating SiO\textsubscript{2} layer and a backside Au mirror, which eliminates the downwards leakage of photons into the substrate. However, the circular symmetry of the nanostructure is broken by introducing a small ellipticity (Figure \ref{Fig1}b), which lifts the degeneracy of the fundamental cavity modes. The resulting mode splitting is proportional to the difference between the major and minor axis of the elliptical cavity \cite{Wang.2019c}.

Figure \ref{Fig1}c shows the two non-degenerate broadband cavity modes of our main EBG device centred at 1279.69 nm (labelled \textit{H}) and 1297.36 nm (labelled \textit{V}), respectively. 
Polarization-resolved measurements (Figure \ref{Fig1}d) confirm that the modes are orthogonally polarized, with degree of linear polarization (DLP) of 79.4\% for the \textit{H} mode and 96.3\% for the \textit{V} mode. We attribute the lower DLP of the \textit{H} mode to small fabrication imperfections along the minor axis of the elliptical cavity.
To demonstrate that the separation of the modes can be easily optimized by varying the ellipticity of the device in the initial design phase, in Figure \ref{Fig1}e we report the central wavelength of the \textit{H} and \textit{V} cavity modes measured on a series of 6 EBGs with fixed radius R\textsubscript{\textit{H}} = 515 nm and variable radius R\textsubscript{\textit{V}}. We observe that a 10 nm change in R\textsubscript{\textit{V}} leads to an 8.4 nm shift of the \textit{V} mode and a 2.3 nm shift of the \textit{H} mode. The latter is caused by variations in the spatial profile of the \textit{H} mode when the ellipticity of the central disk is modified by varying R\textsubscript{\textit{V}}. As predicted by numerical simulations, the resulting mode splitting varies almost linearly with respect to the difference between the major and minor radius and vanishes for a circular device (Figure \ref{Fig1}f).
%
\begin{figure}[h!]
\includegraphics[width=1\textwidth]{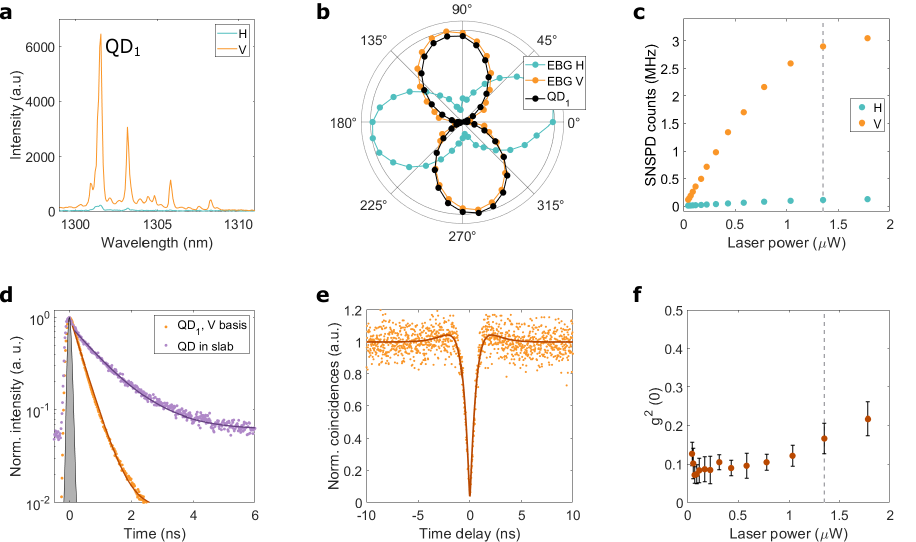}
\caption{\textbf{Characterization of the linearly polarized QD source.} 
(a) Photoluminescence spectrum of a QD embedded in the EBG cavity. The brightest transition at  1301.5 nm is labelled QD\textsubscript{1}. Photons are collected in the \textit{H} and \textit{V} polarizations.
(b) Polarization-resolved photoluminescence intensity of the QD\textsubscript{1} transition, which is aligned with the \textit{V} mode of the EBG and shows a degree of linear polarization of 96.1\%.
(c) SNSPD count rate for the \textit{H} and \textit{V} components of the QD\textsubscript{1} transition as a function of the pump power. The dashed line marks the value of CW pump power chosen as \textit{P\textsubscript{sat}}.
(d) Radiative lifetime of the QD\textsubscript{1} transition ($\sim$ 402 ps) and of a reference QD transition in the unetched GaAs slab ($\sim$ 1.21 ns) measured under pulsed excitation at 80 MHz. The gray area indicates the instrument response function, while the solid lines represent the exponential fit of the decay traces. 
(e) Second order correlation function \textit{g\textsuperscript{2}($\tau$)} of QD\textsubscript{1} measured under a CW pump power of 0.5\textit{P\textsubscript{sat}}. The value of the \textit{g\textsuperscript{2}(0)} extracted from the raw data is 0.104 $\pm$ 0.021. Fitting the HBT data with a four-level model (solid line) yields \textit{g\textsuperscript{2}(0)} = 0.055 $\pm$ 0.034. 
(f) Values of the \textit{g\textsuperscript{2}(0)} extracted from the raw data as a function of the pump power. The dashed line marks the value of \textit{P\textsubscript{sat}}.}
\label{Fig2} 
\end{figure}

When a QD transition is resonant with one of the polarized EBG modes, the cavity channels the spontaneous emission into the \textit{H} and \textit{V} polarizations with a ratio that depends on the mode splitting, the mode linewidth, and the Purcell factor experienced by the QD \cite{Wang.2019c}. In Figure \ref{Fig2}a we report the PL spectrum of a QD coupled to the \textit{V} mode of our main EBG device: despite the broad nature of the bullseye cavity modes, the large splitting allows the emitted photons to be efficiently prepared in the \textit{V} polarization. Polarization-resolved measurements of the integrated PL intensity (Figure \ref{Fig2}b) confirm that the QD\textsubscript{1} transition displays a high DLP of 96.1\% and is perfectly aligned with the \textit{V} mode of the EBG. 
As shown by the power dependence in Figure \ref{Fig2}c, the QD emission saturates for a CW laser power \textit{P\textsubscript{sat}} of approximately 1.3 $\mu$W. After spectral filtering of the QD\textsubscript{1} transition, up to 3 million \textit{V} polarized photon counts per second are detected by a superconducting nanowire single photon detector (SNSPD) with an efficiency of $\sim$ 65\%. For comparison, less than 100 000 photons counts per second are detected in the \textit{H} polarization.

%
%
\begin{figure}[h!]
\includegraphics[width=1\textwidth]{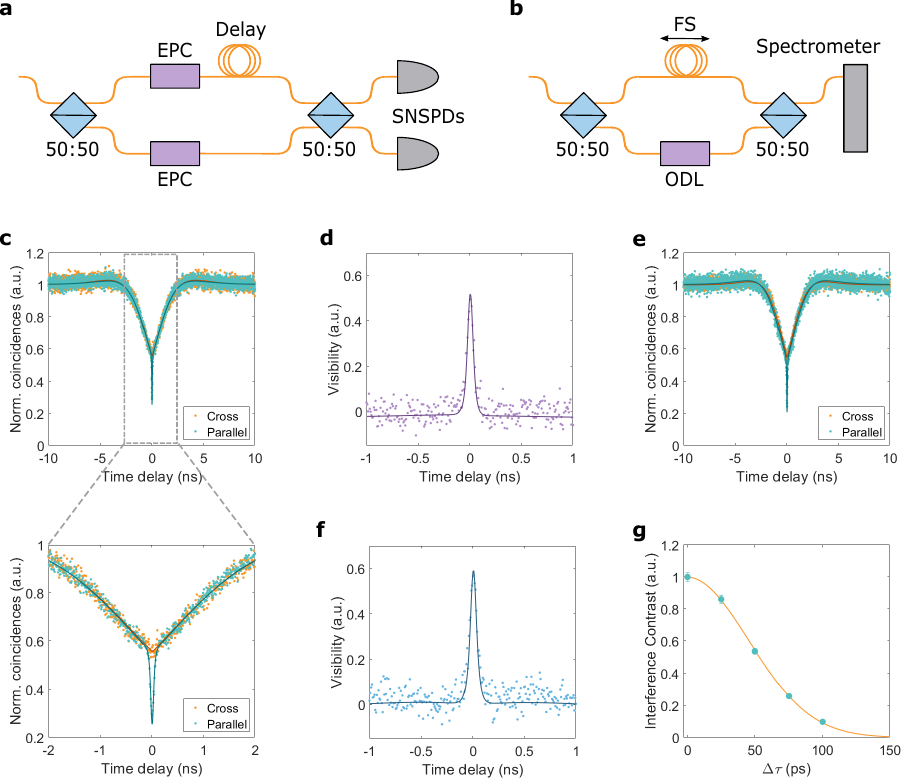}
\caption{\textbf{Characterization of the two-photon interference (TPI).}
(a) Schematic of the HOM experimental setup for TPI measurements. The polarization of photons propagating in the two arms is set by electronic polarization controllers (EPCs). Coincidences are detected by a pair of superconducting nanowire single-photon detectors.
(b) Schematic of the Mach-Zehnder interferometer for coherence time measurements. One of the arms contains a fiber stretcher (FS) to generate interference fringes whereas the other one contains an optical delay line (ODL) to vary the delay between the arms on longer timescales.
(c) Normalized coincidences for distinguishable (orange) and indistinguishable (teal) photons in a TPI experiment performed without any external polarization filtering. The bottom panel shows a detail of the fit on a time scale of $\pm$ 2 ns.
(d) TPI visibility from (c).
(e) Normalized coincidences and visibility for the same TPI experiment repeated after adding a linear polarizer in the collection arm of the microscope, which reduces the the single-photon count rate by less than 5\%.
(f) TPI visibility from (e).
(g) Coherence time of the QD\textsubscript{1} transition measured in a fiber-based Mach-Zehnder interferometer. Fitting the contrast decay with a Voigt profile (orange line) yields a coherence time of 63 $\pm$ 16 ps.
}
\label{Fig3} 
\end{figure}
The Purcell enhancement provided by the EBG cavity is quantified by measuring the radiative lifetimes under above-band pulsed excitation. In Figure \ref{Fig2}d we compare the time-resolved luminescence trace of QD\textsubscript{1}, which decays with a time constant $\tau \textsubscript{QD1} \sim$ 402 ps, to the one of a reference quantum dot in an unpatterned region of the sample. Assuming a typical lifetime $\tau \textsubscript{slab}$ = 1.594 $\pm$ 0.254 ns for QDs in the bare GaAs slab based on our recent work on a similar sample \cite{Barbiero.2022b}, we estimate a Purcell factor \textit{F\textsubscript{p}} = 3.9 $\pm$ 0.6.

The purity of the EBG single-photon source is investigated using a fiber-based Hanbury Brown and Twiss (HBT) setup. As shown in Figure \ref{Fig2}e, the second order correlation function \textit{g\textsuperscript{2}($\tau$)} measured under a CW pump power of 0.5\textit{P\textsubscript{sat}} exhibits the typical antibunching dip at zero delay. The \textit{g\textsuperscript{2}(0)} = 0.055 $\pm$ 0.034 is extracted by fitting the HBT coincidences with a four-level model \cite{Anderson.2020}. The occurrence of multiphoton events is found to increase with the excitation power (Figure \ref{Fig2}f), due to saturation of the QD\textsubscript{1} transition and increased background emission. 

As an example of how useful the strong polarization-selective enhancement provided by the EBG is, we now show two-photon interference (TPI) without any external polarization filtering in a fiber-based Hong-Ou-Mandel \cite{Hong.1987} (HOM) setup (Figure \ref{Fig3}a). 
In this experiment, the expected correlations for cross-polarized ($\perp$) photons are given by \cite{Anderson.2021}:
\begin{equation}
g_{\perp}^{2} (\tau) = \frac{1}{2}g^{2} (\tau) 
+\frac{1}{4} \left[ g^{2} (\tau + \Delta\tau)+ g^{2} (\tau - \Delta\tau)\right],
\label{eq1}
\end{equation}
where $\Delta\tau$ is the delay between photons determined by the setup. The first term describes the probability of detecting coincidences between two photons that travel through the same arm of the interferometer, while the second term describes the expected correlations between photons that travel through different arms.
In the co-polarised ($\parallel$) case we expect an additional interference dip at zero delay due to the HOM effect, which modifies the above equation to give:
\begin{equation}
g_{\parallel}^{2} (\tau) = \frac{1}{2}g^{2} (\tau) 
+\frac{1}{4} \left[ g^{2} (\tau + \Delta\tau)+ g^{2} (\tau - \Delta\tau)\right] (1-e^{-2|\tau| / T_2}),
\label{eq2}
\end{equation}
where $T_2$ is the coherence time of the QD transition.
The resulting two-photon interference visibility in our setup can be calculated as:
\begin{equation}
V_{HOM} = 1 - \frac{g_{\parallel}^{2} (\tau)}{g_{\perp}^{2} (\tau)}.
\label{eq3}
\end{equation}
In Figure \ref{Fig3}c we report the measured correlations for co- and cross-polarized photons emitted by the EBG source without any external polarization filtering. From the contrast between the co- and cross-polarized coincidences at zero delay we extract a maximum raw TPI visibility $V_{HOM}$ = 0.543 $\pm$ 0.023 (Figure \ref{Fig3}d). 
For comparison, Figure \ref{Fig3}e shows the measured correlations when the residual \textit{H} polarized photons are removed using a linear polarizer. Adding external filtering increases the maximum visibility to 0.611 $\pm$ 0.032 (Figure \ref{Fig3}f), while the loss in the single-photon count rate is limited to < 5\% thanks to the high DLP displayed by the EBG source. By fitting the TPI visibility we extract a width of the interference window of 71 ps, which is similar to the coherence time $T_2$ = 63 $\pm$ 16 ps of the QD\textsubscript{1} transition measured directly using a fiber-based Mach-Zehnder interferometer (Figure \ref{Fig3}b and \ref{Fig3}g). These values are comparable to the typical coherence times reported in literature for Stranski-Krastanov QDs emitting in the telecom O-band \cite{Felle.2015, Xiang.2020}.
%
%
\begin{figure}[h]
\includegraphics[width=1\textwidth]{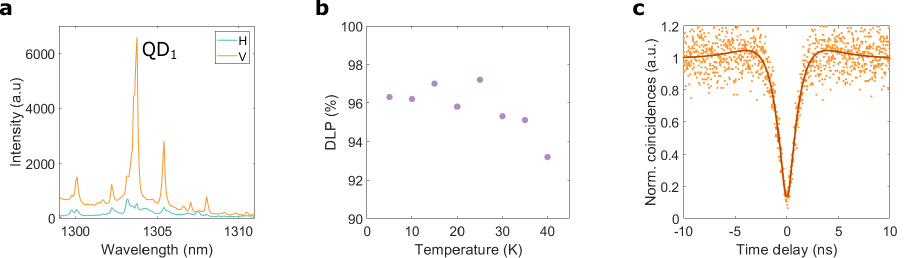}
\caption{\textbf{Performance of the linearly polarized QD source at high temperature.} 
(a) Photoluminescence spectrum of the QD source at a temperature of 40 K. Photons are collected in the \textit{H} and \textit{V} polarizations.
(b) Degree of linear polarization of the QD\textsubscript{1} transition as a function of temperature.
(c) Second order correlation function \textit{g\textsuperscript{2}($\tau$)} of QD\textsubscript{1} measured at a temperature of 40 K under a CW pump power of 0.3\textit{P\textsubscript{sat}}. The value \textit{g\textsuperscript{2}(0)} = 0.137 $\pm$ 0.031 is extracted by fitting the HBT data with a four level model (solid line). 
}
\label{Fig4} 
\end{figure}
%

Finally, it is worth noting that QD single-photon sources operating at relatively high temperatures are most desirable for practical quantum network applications, which require not only  integrability with the fiber infrastructure but also deployment outside the laboratory environment.
Therefore, given the current state-of-the-art in QD operation within compact Stirling cryocoolers \cite{Schlehahn.2015,Musia.2020,Gao.2022}, in Figure \ref{Fig4} we test the performance of our EBG source at temperatures up to 40 K.  
Despite a $\sim$ 3 nm redshift of the QD spectrum (Figure \ref{Fig4}a), the broadband nature of the bullseye cavity keeps the QD\textsubscript{1} transition coupled to the \textit{V} mode. As a result, we observe a high DLP $\ge$ 93\% in the whole range between 5 K and 40 K (Figure \ref{Fig4}b). Operating the source in such a broad temperature range would be very challenging with narrowband devices such as micropillars, which usually require a temperature dependent optimization of the QD-cavity coupling \cite{Liu.2017b}.
Moreover, the second order correlation function \textit{g\textsuperscript{2}($\tau$)} in Figure \ref{Fig4}c confirms that the purity of the source has not substantially deteriorated, with \textit{g\textsuperscript{2}(0)} = 0.137 $\pm$ 0.031.

\section{Summary}
\noindent We reported on polarization-selective enhancement of an InAs/GaAs QD emitting in the telecom O-band coupled to an elliptical Bragg grating.
We designed the device to achieve a large separation between the non-degenerate cavity modes such that, thanks to a mode splitting-to-linewidth ratio close to 2, the QD emits preferentially in the \textit{V} polarized mode.
We took advantage of this property to demonstrate bright emission of linearly polarized single photons under above-band excitation with DLP > 96\% and Purcell factor of 3.9.
Furthermore, we presented a measurement of two-photon interference without any external polarization filtering and showed that the source can be operated in a broad temperature range up to 40 K with minimal impact on its polarization-selective properties, which facilitates the integration in compact modules for field deployment.

In future experiments, the \textit{H} cavity mode could be exploited to resonantly excite the QD without any polarization-based laser suppression \cite{Wang.2019c} and reveal an enhanced quality of the source in terms of purity and indistinguishability.
Moreover, one could take advantage of the broadband selective enhancement provided by the EBG  to generate cycling transitions without the need for QD tunability.
Finally, it is worth noting that the polarized single-photon efficiency strongly depends on the Purcell factor provided by the microcavity. Higher values of \textit{F\textsubscript{p}} could be achieved by employing in-situ electron beam lithography for an optimal spatial and spectral coupling between the QD emitter and the cavity mode.
%

%
\section{Methods}
\subsection{Sample growth and device fabrication}
The QD sample consists of an undoped (100) GaAs substrate, a 200 nm thick Al\textsubscript{0.8}Ga\textsubscript{0.2}As sacrificial layer, and a  240 nm thick GaAs membrane containing SK InAs QDs. 
After the epitaxial growth, 300 nm of SiO\textsubscript{2} is deposited on the surface of the GaAs slab, followed by approximately 200 nm of Au. 
The epitaxial wafer is then bonded to a GaAs carrier, and the original substrate and Al\textsubscript{0.8}Ga\textsubscript{0.2}As sacrificial layer are removed using the membrane-transfer technique presented in our recent work \cite{Barbiero.2022b}. 
Finally, the EBGs are non-deterministically fabricated on the GaAs membrane using electron-beam lithography and a chlorine-based dry etch process.
The thickness of the GaAs and SiO\textsubscript{2} layers, and the design parameters of the EBGs are optimized using FEM numerical simulations in the frequency domain.

\subsection{Optical characterization}
The devices are operated at a temperature of 5 K in a closed-cycle cryostat and  optically pumped using either a CW or a pulsed laser diode (80 MHz) at a wavelength of 785 nm. QD emission from a single device is collected by a fiber-coupled confocal microscope with a NA = 0.5 objective lens. For polarization-resolved measurements, a rotating half-wave plate and a fixed linear polarizer are mounted in the collection arm of the microscope.
The collected photons are then sent to either a spectrometer equipped with an InGaAs photodiode array for spectal analysis, or to a tunable optical filter with 0.25 nm FWHM bandwidth in order to isolate a single transition. After spectral filtering, photons are guided to either a fiber-based HBT setup or to a HOM interferometer with electronic polarization controllers (EPCs) placed in each arm \cite{Anderson.2021}. Finally, photons are detected detected using a pair of superconducting nanowire single-photon detectors with $\sim$40 ps jitter.

\section{Author contributions}
R.M.S, A.J.S and D.A.R. guided and supervised the project. L.E.G. engineered the epitaxial process and grew the sample. A.B. optimized the device design using FEM simulations. A.B. and J.S.-S. fabricated the devices. A.B. and G.S. performed the measurements. A.B.  and T.M analysed  the data. A.B. wrote the manuscript. All authors discussed the results and commented on the manuscript. The authors declare that they have no competing financial interests.

\begin{acknowledgement}
\noindent The authors thank D. Ellis and B. Ramsay for the PECVD SiO\textsubscript{2} deposition. They acknowledge funding from the Ministry of Internal Affairs and Communications, Japan, via the project of ICT priority technology (JPMI00316) ‘Research and Development for Construction of a Global Quantum Cryptography Network’. They further acknowledge funding from QFoundry (project number 48484), which is part-funded by the UK Quantum Technologies Challenge under UK Research and Innovation (UKRI).
\end{acknowledgement}


\bibliography{Oband_EBG_Arxiv}

\providecommand{\latin}[1]{#1}
\makeatletter
\providecommand{\doi}
  {\begingroup\let\do\@makeother\dospecials
  \catcode`\{=1 \catcode`\}=2 \doi@aux}
\providecommand{\doi@aux}[1]{\endgroup\texttt{#1}}
\makeatother
\providecommand*\mcitethebibliography{\thebibliography}
\csname @ifundefined\endcsname{endmcitethebibliography}
  {\let\endmcitethebibliography\endthebibliography}{}
\begin{mcitethebibliography}{46}
\providecommand*\natexlab[1]{#1}
\providecommand*\mciteSetBstSublistMode[1]{}
\providecommand*\mciteSetBstMaxWidthForm[2]{}
\providecommand*\mciteBstWouldAddEndPuncttrue
  {\def\EndOfBibitem{\unskip.}}
\providecommand*\mciteBstWouldAddEndPunctfalse
  {\let\EndOfBibitem\relax}
\providecommand*\mciteSetBstMidEndSepPunct[3]{}
\providecommand*\mciteSetBstSublistLabelBeginEnd[3]{}
\providecommand*\EndOfBibitem{}
\mciteSetBstSublistMode{f}
\mciteSetBstMaxWidthForm{subitem}{(\alph{mcitesubitemcount})}
\mciteSetBstSublistLabelBeginEnd
  {\mcitemaxwidthsubitemform\space}
  {\relax}
  {\relax}

\bibitem[Gisin \latin{et~al.}(2002)Gisin, Ribordy, Tittel, and
  Zbinden]{Gisin.2002}
Gisin,~N.; Ribordy,~G.; Tittel,~W.; Zbinden,~H. {Quantum cryptography}.
  \emph{{Reviews of Modern Physics}} \textbf{2002}, \emph{74}, 145--195\relax
\mciteBstWouldAddEndPuncttrue
\mciteSetBstMidEndSepPunct{\mcitedefaultmidpunct}
{\mcitedefaultendpunct}{\mcitedefaultseppunct}\relax
\EndOfBibitem
\bibitem[Xu \latin{et~al.}(2020)Xu, Ma, Zhang, Lo, and Pan]{Xu.2020}
Xu,~F.; Ma,~X.; Zhang,~Q.; Lo,~H.-K.; Pan,~J.-W. {Secure quantum key
  distribution with realistic devices}. \emph{{Reviews of Modern Physics}}
  \textbf{2020}, \emph{92}, 131\relax
\mciteBstWouldAddEndPuncttrue
\mciteSetBstMidEndSepPunct{\mcitedefaultmidpunct}
{\mcitedefaultendpunct}{\mcitedefaultseppunct}\relax
\EndOfBibitem
\bibitem[Couteau \latin{et~al.}(2023)Couteau, Barz, Durt, Gerrits, Huwer,
  Prevedel, Rarity, Shields, and Weihs]{Couteau.2023}
Couteau,~C.; Barz,~S.; Durt,~T.; Gerrits,~T.; Huwer,~J.; Prevedel,~R.;
  Rarity,~J.; Shields,~A.; Weihs,~G. {Applications of single photons to quantum
  communication and computing}. \emph{{Nature Reviews Physics}} \textbf{2023},
  \emph{5}, 326--338\relax
\mciteBstWouldAddEndPuncttrue
\mciteSetBstMidEndSepPunct{\mcitedefaultmidpunct}
{\mcitedefaultendpunct}{\mcitedefaultseppunct}\relax
\EndOfBibitem
\bibitem[Couteau \latin{et~al.}(2023)Couteau, Barz, Durt, Gerrits, Huwer,
  Prevedel, Rarity, Shields, and Weihs]{Couteau.2023b}
Couteau,~C.; Barz,~S.; Durt,~T.; Gerrits,~T.; Huwer,~J.; Prevedel,~R.;
  Rarity,~J.; Shields,~A.; Weihs,~G. {Applications of single photons in quantum
  metrology, biology and the foundations of quantum physics}. \emph{{Nature
  Reviews Physics}} \textbf{2023}, \emph{5}, 354--363\relax
\mciteBstWouldAddEndPuncttrue
\mciteSetBstMidEndSepPunct{\mcitedefaultmidpunct}
{\mcitedefaultendpunct}{\mcitedefaultseppunct}\relax
\EndOfBibitem
\bibitem[Kok \latin{et~al.}(2007)Kok, Munro, Nemoto, Ralph, Dowling, and
  Milburn]{Kok.2007}
Kok,~P.; Munro,~W.~J.; Nemoto,~K.; Ralph,~T.~C.; Dowling,~J.~P.; Milburn,~G.~J.
  {Linear optical quantum computing with photonic qubits}. \emph{{Physical
  review A, Atomic, molecular, and optical physics}} \textbf{2007}, \emph{79},
  135--174\relax
\mciteBstWouldAddEndPuncttrue
\mciteSetBstMidEndSepPunct{\mcitedefaultmidpunct}
{\mcitedefaultendpunct}{\mcitedefaultseppunct}\relax
\EndOfBibitem
\bibitem[Slussarenko and Pryde(2019)Slussarenko, and Pryde]{Slussarenko.2019}
Slussarenko,~S.; Pryde,~G.~J. {Photonic quantum information processing: A
  concise review}. \emph{{Applied Physics Reviews}} \textbf{2019}, \emph{6},
  041303\relax
\mciteBstWouldAddEndPuncttrue
\mciteSetBstMidEndSepPunct{\mcitedefaultmidpunct}
{\mcitedefaultendpunct}{\mcitedefaultseppunct}\relax
\EndOfBibitem
\bibitem[Thomas \latin{et~al.}(2021)Thomas, Billard, Coste, Wein, Priya,
  Ollivier, Krebs, Taza{\"i}rt, Harouri, Lemaitre, Sagnes, Anton, Lanco,
  Somaschi, Loredo, and Senellart]{Thomas.2021b}
Thomas,~S.~E. \latin{et~al.}  {Bright Polarized Single-Photon Source Based on a
  Linear Dipole}. \emph{{Physical review letters}} \textbf{2021}, \emph{126},
  233601\relax
\mciteBstWouldAddEndPuncttrue
\mciteSetBstMidEndSepPunct{\mcitedefaultmidpunct}
{\mcitedefaultendpunct}{\mcitedefaultseppunct}\relax
\EndOfBibitem
\bibitem[Shields(2007)]{Shields.2007}
Shields,~A.~J. {Semiconductor quantum light sources}. \emph{{Nature Photonics}}
  \textbf{2007}, \emph{1}, 215--223\relax
\mciteBstWouldAddEndPuncttrue
\mciteSetBstMidEndSepPunct{\mcitedefaultmidpunct}
{\mcitedefaultendpunct}{\mcitedefaultseppunct}\relax
\EndOfBibitem
\bibitem[He \latin{et~al.}(2013)He, He, Wei, Wu, Atat{\"u}re, Schneider,
  H{\"o}fling, Kamp, Lu, and Pan]{He.2013}
He,~Y.-M.; He,~Y.; Wei,~Y.-J.; Wu,~D.; Atat{\"u}re,~M.; Schneider,~C.;
  H{\"o}fling,~S.; Kamp,~M.; Lu,~C.-Y.; Pan,~J.-W. {On-demand semiconductor
  single-photon source with near-unity indistinguishability}. \emph{{Nature
  Nanotechnology}} \textbf{2013}, \emph{8}, 213--217\relax
\mciteBstWouldAddEndPuncttrue
\mciteSetBstMidEndSepPunct{\mcitedefaultmidpunct}
{\mcitedefaultendpunct}{\mcitedefaultseppunct}\relax
\EndOfBibitem
\bibitem[Ding \latin{et~al.}(2016)Ding, He, Duan, Gregersen, Chen, Unsleber,
  Maier, Schneider, Kamp, H{\"o}fling, Lu, and Pan]{Ding.2016}
Ding,~X.; He,~Y.; Duan,~Z.-C.; Gregersen,~N.; Chen,~M.-C.; Unsleber,~S.;
  Maier,~S.; Schneider,~C.; Kamp,~M.; H{\"o}fling,~S.; Lu,~C.-Y.; Pan,~J.-W.
  {On-Demand Single Photons with High Extraction Efficiency and Near-Unity
  Indistinguishability from a Resonantly Driven Quantum Dot in a Micropillar}.
  \emph{{Physical review letters}} \textbf{2016}, \emph{116}, 020401\relax
\mciteBstWouldAddEndPuncttrue
\mciteSetBstMidEndSepPunct{\mcitedefaultmidpunct}
{\mcitedefaultendpunct}{\mcitedefaultseppunct}\relax
\EndOfBibitem
\bibitem[Senellart \latin{et~al.}(2017)Senellart, Solomon, and
  White]{Senellart.2017}
Senellart,~P.; Solomon,~G.; White,~A. {High-performance semiconductor
  quantum-dot single-photon sources}. \emph{{Nature Nanotechnology}}
  \textbf{2017}, \emph{12}, 1026--1039\relax
\mciteBstWouldAddEndPuncttrue
\mciteSetBstMidEndSepPunct{\mcitedefaultmidpunct}
{\mcitedefaultendpunct}{\mcitedefaultseppunct}\relax
\EndOfBibitem
\bibitem[Arakawa and Holmes(2020)Arakawa, and Holmes]{Arakawa.2020}
Arakawa,~Y.; Holmes,~M.~J. {Progress in quantum-dot single photon sources for
  quantum information technologies: A broad spectrum overview}. \emph{{Applied
  Physics Reviews}} \textbf{2020}, \emph{7}, 021309\relax
\mciteBstWouldAddEndPuncttrue
\mciteSetBstMidEndSepPunct{\mcitedefaultmidpunct}
{\mcitedefaultendpunct}{\mcitedefaultseppunct}\relax
\EndOfBibitem
\bibitem[Zinoni \latin{et~al.}(2006)Zinoni, Alloing, Monat, Zwiller, Li, Fiore,
  Lunghi, Gerardino, de~Riedmatten, Zbinden, and Gisin]{Zinoni.2006}
Zinoni,~C.; Alloing,~B.; Monat,~C.; Zwiller,~V.; Li,~L.~H.; Fiore,~A.;
  Lunghi,~L.; Gerardino,~A.; de~Riedmatten,~H.; Zbinden,~H.; Gisin,~N.
  {Time-resolved and antibunching experiments on single quantum dots at
  1300nm}. \emph{{Applied Physics Letters}} \textbf{2006}, \emph{88}, 691\relax
\mciteBstWouldAddEndPuncttrue
\mciteSetBstMidEndSepPunct{\mcitedefaultmidpunct}
{\mcitedefaultendpunct}{\mcitedefaultseppunct}\relax
\EndOfBibitem
\bibitem[Ward \latin{et~al.}(2014)Ward, Dean, Stevenson, Bennett, Ellis,
  Cooper, Farrer, Nicoll, Ritchie, and Shields]{Ward.2014}
Ward,~M.~B.; Dean,~M.~C.; Stevenson,~R.~M.; Bennett,~A.~J.; Ellis,~D.;
  Cooper,~K.; Farrer,~I.; Nicoll,~C.~A.; Ritchie,~D.~A.; Shields,~A.~J.
  {Coherent dynamics of a telecom-wavelength entangled photon source}.
  \emph{{Nature communications}} \textbf{2014}, \emph{5}, 3316\relax
\mciteBstWouldAddEndPuncttrue
\mciteSetBstMidEndSepPunct{\mcitedefaultmidpunct}
{\mcitedefaultendpunct}{\mcitedefaultseppunct}\relax
\EndOfBibitem
\bibitem[Portalupi \latin{et~al.}(2019)Portalupi, Jetter, and
  Michler]{Portalupi.2019}
Portalupi,~S.~L.; Jetter,~M.; Michler,~P. {InAs quantum dots grown on
  metamorphic buffers as non-classical light sources at telecom C-band: a
  review}. \emph{{Semiconductor Science and Technology}} \textbf{2019},
  \emph{34}, 053001\relax
\mciteBstWouldAddEndPuncttrue
\mciteSetBstMidEndSepPunct{\mcitedefaultmidpunct}
{\mcitedefaultendpunct}{\mcitedefaultseppunct}\relax
\EndOfBibitem
\bibitem[Anderson \latin{et~al.}(2021)Anderson, M{\"u}ller, Skiba-Szymanska,
  Krysa, Huwer, Stevenson, Heffernan, Ritchie, and Shields]{Anderson.2021}
Anderson,~M.; M{\"u}ller,~T.; Skiba-Szymanska,~J.; Krysa,~A.~B.; Huwer,~J.;
  Stevenson,~R.~M.; Heffernan,~J.; Ritchie,~D.~A.; Shields,~A.~J. {Coherence in
  single photon emission from droplet epitaxy and Stranski--Krastanov quantum
  dots in the telecom C-band}. \emph{{Applied Physics Letters}} \textbf{2021},
  \emph{118}, 014003\relax
\mciteBstWouldAddEndPuncttrue
\mciteSetBstMidEndSepPunct{\mcitedefaultmidpunct}
{\mcitedefaultendpunct}{\mcitedefaultseppunct}\relax
\EndOfBibitem
\bibitem[Kimble(2008)]{Kimble.2008}
Kimble,~H.~J. {The quantum internet}. \emph{{Nature}} \textbf{2008},
  \emph{453}, 1023--1030\relax
\mciteBstWouldAddEndPuncttrue
\mciteSetBstMidEndSepPunct{\mcitedefaultmidpunct}
{\mcitedefaultendpunct}{\mcitedefaultseppunct}\relax
\EndOfBibitem
\bibitem[Somaschi \latin{et~al.}(2016)Somaschi, Giesz, de~Santis, Loredo,
  Almeida, Hornecker, Portalupi, Grange, Ant{\'o}n, Demory, G{\'o}mez, Sagnes,
  Lanzillotti-Kimura, Lema{\'i}tre, Auffeves, White, Lanco, and
  Senellart]{Somaschi.2016}
Somaschi,~N. \latin{et~al.}  {Near-optimal single-photon sources in the solid
  state}. \emph{{Nature Photonics}} \textbf{2016}, \emph{10}, 340--345\relax
\mciteBstWouldAddEndPuncttrue
\mciteSetBstMidEndSepPunct{\mcitedefaultmidpunct}
{\mcitedefaultendpunct}{\mcitedefaultseppunct}\relax
\EndOfBibitem
\bibitem[Gregersen \latin{et~al.}(2016)Gregersen, McCutcheon, M{\o}rk,
  G{\'e}rard, and Claudon]{Gregersen.2016}
Gregersen,~N.; McCutcheon,~D. P.~S.; M{\o}rk,~J.; G{\'e}rard,~J.-M.;
  Claudon,~J. {A broadband tapered nanocavity for efficient nonclassical light
  emission}. \emph{{Optics express}} \textbf{2016}, \emph{24},
  20904--20924\relax
\mciteBstWouldAddEndPuncttrue
\mciteSetBstMidEndSepPunct{\mcitedefaultmidpunct}
{\mcitedefaultendpunct}{\mcitedefaultseppunct}\relax
\EndOfBibitem
\bibitem[Liu \latin{et~al.}(2018)Liu, Brash, O'Hara, Martins, Phillips, Coles,
  Royall, Clarke, Bentham, Prtljaga, Itskevich, Wilson, Skolnick, and
  Fox]{Liu.2018}
Liu,~F.; Brash,~A.~J.; O'Hara,~J.; Martins,~L. M. P.~P.; Phillips,~C.~L.;
  Coles,~R.~J.; Royall,~B.; Clarke,~E.; Bentham,~C.; Prtljaga,~N.;
  Itskevich,~I.~E.; Wilson,~L.~R.; Skolnick,~M.~S.; Fox,~A.~M. {High Purcell
  factor generation of indistinguishable on-chip single photons}. \emph{{Nature
  Nanotechnology}} \textbf{2018}, \emph{13}, 835--840\relax
\mciteBstWouldAddEndPuncttrue
\mciteSetBstMidEndSepPunct{\mcitedefaultmidpunct}
{\mcitedefaultendpunct}{\mcitedefaultseppunct}\relax
\EndOfBibitem
\bibitem[Uppu \latin{et~al.}(2020)Uppu, Pedersen, Wang, Olesen, Papon, Zhou,
  Midolo, Scholz, Wieck, Ludwig, and Lodahl]{Uppu.2020}
Uppu,~R.; Pedersen,~F.~T.; Wang,~Y.; Olesen,~C.~T.; Papon,~C.; Zhou,~X.;
  Midolo,~L.; Scholz,~S.; Wieck,~A.~D.; Ludwig,~A.; Lodahl,~P. {Scalable
  integrated single-photon source}. \emph{{Science advances}} \textbf{2020},
  \emph{6}\relax
\mciteBstWouldAddEndPuncttrue
\mciteSetBstMidEndSepPunct{\mcitedefaultmidpunct}
{\mcitedefaultendpunct}{\mcitedefaultseppunct}\relax
\EndOfBibitem
\bibitem[Tomm \latin{et~al.}(2021)Tomm, Javadi, Antoniadis, Najer, L{\"o}bl,
  Korsch, Schott, Valentin, Wieck, Ludwig, and Warburton]{Tomm.2021}
Tomm,~N.; Javadi,~A.; Antoniadis,~N.~O.; Najer,~D.; L{\"o}bl,~M.~C.;
  Korsch,~A.~R.; Schott,~R.; Valentin,~S.~R.; Wieck,~A.~D.; Ludwig,~A.;
  Warburton,~R.~J. {A bright and fast source of coherent single photons}.
  \emph{{Nature Nanotechnology}} \textbf{2021}, \emph{16}, 399--403\relax
\mciteBstWouldAddEndPuncttrue
\mciteSetBstMidEndSepPunct{\mcitedefaultmidpunct}
{\mcitedefaultendpunct}{\mcitedefaultseppunct}\relax
\EndOfBibitem
\bibitem[Davan{\c{c}}o \latin{et~al.}(2011)Davan{\c{c}}o, Rakher, Schuh,
  Badolato, and Srinivasan]{Davanco.2011}
Davan{\c{c}}o,~M.; Rakher,~M.~T.; Schuh,~D.; Badolato,~A.; Srinivasan,~K. {A
  circular dielectric grating for vertical extraction of single quantum dot
  emission}. \emph{{Applied Physics Letters}} \textbf{2011}, \emph{99},
  041102\relax
\mciteBstWouldAddEndPuncttrue
\mciteSetBstMidEndSepPunct{\mcitedefaultmidpunct}
{\mcitedefaultendpunct}{\mcitedefaultseppunct}\relax
\EndOfBibitem
\bibitem[Liu \latin{et~al.}(2019)Liu, Su, Wei, Yao, Silva, Yu, Iles-Smith,
  Srinivasan, Rastelli, Li, and Wang]{Liu.2019}
Liu,~J.; Su,~R.; Wei,~Y.; Yao,~B.; Silva,~S. F. C.~d.; Yu,~Y.; Iles-Smith,~J.;
  Srinivasan,~K.; Rastelli,~A.; Li,~J.; Wang,~X. {A solid-state source of
  strongly entangled photon pairs with high brightness and
  indistinguishability}. \emph{{Nature Nanotechnology}} \textbf{2019},
  \emph{14}, 586--593\relax
\mciteBstWouldAddEndPuncttrue
\mciteSetBstMidEndSepPunct{\mcitedefaultmidpunct}
{\mcitedefaultendpunct}{\mcitedefaultseppunct}\relax
\EndOfBibitem
\bibitem[Wang \latin{et~al.}(2019)Wang, Hu, Chung, Qin, Yang, Li, Liu, Zhong,
  He, Ding, Deng, Dai, Huo, H{\"o}fling, Lu, and Pan]{Wang.2019}
Wang,~H. \latin{et~al.}  {On-Demand Semiconductor Source of Entangled Photons
  Which Simultaneously Has High Fidelity, Efficiency, and
  Indistinguishability}. \emph{{Physical review letters}} \textbf{2019},
  \emph{122}, 113602\relax
\mciteBstWouldAddEndPuncttrue
\mciteSetBstMidEndSepPunct{\mcitedefaultmidpunct}
{\mcitedefaultendpunct}{\mcitedefaultseppunct}\relax
\EndOfBibitem
\bibitem[Rickert \latin{et~al.}(2019)Rickert, Kupko, Rodt, Reitzenstein, and
  Heindel]{Rickert.2019}
Rickert,~L.; Kupko,~T.; Rodt,~S.; Reitzenstein,~S.; Heindel,~T. {Optimized
  designs for telecom-wavelength quantum light sources based on hybrid circular
  Bragg gratings}. \emph{{Optics express}} \textbf{2019}, \emph{27},
  36824\relax
\mciteBstWouldAddEndPuncttrue
\mciteSetBstMidEndSepPunct{\mcitedefaultmidpunct}
{\mcitedefaultendpunct}{\mcitedefaultseppunct}\relax
\EndOfBibitem
\bibitem[Barbiero \latin{et~al.}(2022)Barbiero, Huwer, Skiba-Szymanska,
  M{\"u}ller, Stevenson, and Shields]{Barbiero.2022}
Barbiero,~A.; Huwer,~J.; Skiba-Szymanska,~J.; M{\"u}ller,~T.; Stevenson,~R.~M.;
  Shields,~A.~J. {Design study for an efficient semiconductor quantum light
  source operating in the telecom C-band based on an electrically-driven
  circular Bragg grating}. \emph{{Optics express}} \textbf{2022}, \emph{30},
  10919\relax
\mciteBstWouldAddEndPuncttrue
\mciteSetBstMidEndSepPunct{\mcitedefaultmidpunct}
{\mcitedefaultendpunct}{\mcitedefaultseppunct}\relax
\EndOfBibitem
\bibitem[Bremer \latin{et~al.}(2022)Bremer, Jimenez, Thiele, Weber, Huber,
  Rodt, Herkommer, Burger, H{\"o}fling, Giessen, and Reitzenstein]{Bremer.2022}
Bremer,~L.; Jimenez,~C.; Thiele,~S.; Weber,~K.; Huber,~T.; Rodt,~S.;
  Herkommer,~A.; Burger,~S.; H{\"o}fling,~S.; Giessen,~H.; Reitzenstein,~S.
  {Numerical optimization of single-mode fiber-coupled single-photon sources
  based on semiconductor quantum dots}. \emph{{Optics express}} \textbf{2022},
  \emph{30}, 15913\relax
\mciteBstWouldAddEndPuncttrue
\mciteSetBstMidEndSepPunct{\mcitedefaultmidpunct}
{\mcitedefaultendpunct}{\mcitedefaultseppunct}\relax
\EndOfBibitem
\bibitem[Kolatschek \latin{et~al.}(2021)Kolatschek, Nawrath, Bauer, Huang,
  Fischer, Sittig, Jetter, Portalupi, and Michler]{Kolatschek.2021}
Kolatschek,~S.; Nawrath,~C.; Bauer,~S.; Huang,~J.; Fischer,~J.; Sittig,~R.;
  Jetter,~M.; Portalupi,~S.~L.; Michler,~P. {Bright Purcell Enhanced
  Single-Photon Source in the Telecom O-Band Based on a Quantum Dot in a
  Circular Bragg Grating}. \emph{{Nano letters}} \textbf{2021}, \emph{21},
  7740--7745\relax
\mciteBstWouldAddEndPuncttrue
\mciteSetBstMidEndSepPunct{\mcitedefaultmidpunct}
{\mcitedefaultendpunct}{\mcitedefaultseppunct}\relax
\EndOfBibitem
\bibitem[Barbiero \latin{et~al.}(2022)Barbiero, Huwer, Skiba-Szymanska, Ellis,
  Stevenson, M{\"u}ller, Shooter, Goff, Ritchie, and Shields]{Barbiero.2022b}
Barbiero,~A.; Huwer,~J.; Skiba-Szymanska,~J.; Ellis,~D. J.~P.;
  Stevenson,~R.~M.; M{\"u}ller,~T.; Shooter,~G.; Goff,~L.~E.; Ritchie,~D.~A.;
  Shields,~A.~J. {High-Performance Single-Photon Sources at Telecom Wavelength
  Based on Broadband Hybrid Circular Bragg Gratings}. \emph{{ACS Photonics}}
  \textbf{2022}, \emph{9}, 3060--3066\relax
\mciteBstWouldAddEndPuncttrue
\mciteSetBstMidEndSepPunct{\mcitedefaultmidpunct}
{\mcitedefaultendpunct}{\mcitedefaultseppunct}\relax
\EndOfBibitem
\bibitem[Nawrath \latin{et~al.}(2023)Nawrath, Joos, Kolatschek, Bauer, Pruy,
  Hornung, Fischer, Huang, Vijayan, Sittig, Jetter, Portalupi, and
  Michler]{Nawrath.2023}
Nawrath,~C.; Joos,~R.; Kolatschek,~S.; Bauer,~S.; Pruy,~P.; Hornung,~F.;
  Fischer,~J.; Huang,~J.; Vijayan,~P.; Sittig,~R.; Jetter,~M.;
  Portalupi,~S.~L.; Michler,~P. {Bright Source of Purcell--Enhanced, Triggered,
  Single Photons in the Telecom C--Band}. \emph{{Advanced Quantum
  Technologies}} \textbf{2023}, \emph{5}, 104\relax
\mciteBstWouldAddEndPuncttrue
\mciteSetBstMidEndSepPunct{\mcitedefaultmidpunct}
{\mcitedefaultendpunct}{\mcitedefaultseppunct}\relax
\EndOfBibitem
\bibitem[Unitt \latin{et~al.}(2005)Unitt, Bennett, Atkinson, Ritchie, and
  Shields]{Unitt.2005}
Unitt,~D.~C.; Bennett,~A.~J.; Atkinson,~P.; Ritchie,~D.~A.; Shields,~A.~J.
  {Polarization control of quantum dot single-photon sources via a
  dipole-dependent Purcell effect}. \emph{{Physical Review}} \textbf{2005},
  \emph{72}, 681\relax
\mciteBstWouldAddEndPuncttrue
\mciteSetBstMidEndSepPunct{\mcitedefaultmidpunct}
{\mcitedefaultendpunct}{\mcitedefaultseppunct}\relax
\EndOfBibitem
\bibitem[Lee and Lin(2014)Lee, and Lin]{Lee.2014}
Lee,~Y.-S.; Lin,~S.-D. {Polarized emission of quantum dots in microcavity and
  anisotropic Purcell factors}. \emph{{Optics express}} \textbf{2014},
  \emph{22}, 1512--1523\relax
\mciteBstWouldAddEndPuncttrue
\mciteSetBstMidEndSepPunct{\mcitedefaultmidpunct}
{\mcitedefaultendpunct}{\mcitedefaultseppunct}\relax
\EndOfBibitem
\bibitem[Lee \latin{et~al.}(2019)Lee, Villa, Bennett, Stevenson, Ellis, Farrer,
  Ritchie, and Shields]{Lee.2019}
Lee,~J.~P.; Villa,~B.; Bennett,~A.~J.; Stevenson,~R.~M.; Ellis,~D. J.~P.;
  Farrer,~I.; Ritchie,~D.~A.; Shields,~A.~J. {A quantum dot as a source of
  time-bin entangled multi-photon states}. \emph{{Quantum Science and
  Technology}} \textbf{2019}, \emph{4}, 025011\relax
\mciteBstWouldAddEndPuncttrue
\mciteSetBstMidEndSepPunct{\mcitedefaultmidpunct}
{\mcitedefaultendpunct}{\mcitedefaultseppunct}\relax
\EndOfBibitem
\bibitem[Appel \latin{et~al.}(2021)Appel, Tiranov, Javadi, L{\"o}bl, Wang,
  Scholz, Wieck, Ludwig, Warburton, and Lodahl]{Appel.2021}
Appel,~M.~H.; Tiranov,~A.; Javadi,~A.; L{\"o}bl,~M.~C.; Wang,~Y.; Scholz,~S.;
  Wieck,~A.~D.; Ludwig,~A.; Warburton,~R.~J.; Lodahl,~P. {Coherent Spin-Photon
  Interface with Waveguide Induced Cycling Transitions}. \emph{{Physical review
  letters}} \textbf{2021}, \emph{126}, 013602\relax
\mciteBstWouldAddEndPuncttrue
\mciteSetBstMidEndSepPunct{\mcitedefaultmidpunct}
{\mcitedefaultendpunct}{\mcitedefaultseppunct}\relax
\EndOfBibitem
\bibitem[Peniakov \latin{et~al.}()Peniakov, Buchinger, Helal, Betzold, Reum,
  Rota, Ronco, Beccaceci, Krieger, {Da Silva, Saimon F. Covre}, Rastelli,
  Trotta, Pfenning, Hoefling, and Huber-Loyola]{Peniakov.2023}
Peniakov,~G.; Buchinger,~Q.; Helal,~M.; Betzold,~S.; Reum,~Y.; Rota,~M.~B.;
  Ronco,~G.; Beccaceci,~M.; Krieger,~T.~M.; {Da Silva, Saimon F. Covre},;
  Rastelli,~A.; Trotta,~R.; Pfenning,~A.; Hoefling,~S.; Huber-Loyola,~T.
  {Polarized and Un-Polarized Emission from a Single Emitter in a Bullseye
  Resonator}. \emph{arXiv:2308.06231} \relax
\mciteBstWouldAddEndPunctfalse
\mciteSetBstMidEndSepPunct{\mcitedefaultmidpunct}
{}{\mcitedefaultseppunct}\relax
\EndOfBibitem
\bibitem[Wang \latin{et~al.}(2019)Wang, He, Chung, Hu, Yu, Chen, Ding, Chen,
  Qin, Yang, Liu, Duan, Li, Gerhardt, Winkler, Jurkat, Wang, Gregersen, Huo,
  Dai, Yu, H{\"o}fling, Lu, and Pan]{Wang.2019c}
Wang,~H. \latin{et~al.}  {Towards optimal single-photon sources from polarized
  microcavities}. \emph{{Nature Photonics}} \textbf{2019}, \emph{13},
  770--775\relax
\mciteBstWouldAddEndPuncttrue
\mciteSetBstMidEndSepPunct{\mcitedefaultmidpunct}
{\mcitedefaultendpunct}{\mcitedefaultseppunct}\relax
\EndOfBibitem
\bibitem[Anderson \latin{et~al.}(2020)Anderson, M{\"u}ller, Huwer,
  Skiba-Szymanska, Krysa, Stevenson, Heffernan, Ritchie, and
  Shields]{Anderson.2020}
Anderson,~M.; M{\"u}ller,~T.; Huwer,~J.; Skiba-Szymanska,~J.; Krysa,~A.~B.;
  Stevenson,~R.~M.; Heffernan,~J.; Ritchie,~D.~A.; Shields,~A.~J. {Quantum
  teleportation using highly coherent emission from telecom C-band quantum
  dots}. \emph{{npj Quantum Information}} \textbf{2020}, \emph{6}, 14\relax
\mciteBstWouldAddEndPuncttrue
\mciteSetBstMidEndSepPunct{\mcitedefaultmidpunct}
{\mcitedefaultendpunct}{\mcitedefaultseppunct}\relax
\EndOfBibitem
\bibitem[Hong \latin{et~al.}(1987)Hong, Ou, and Mandel]{Hong.1987}
Hong,~C.~K.; Ou,~Z.~Y.; Mandel,~L. {Measurement of subpicosecond time intervals
  between two photons by interference}. \emph{{Physical review letters}}
  \textbf{1987}, \emph{59}, 2044--2046\relax
\mciteBstWouldAddEndPuncttrue
\mciteSetBstMidEndSepPunct{\mcitedefaultmidpunct}
{\mcitedefaultendpunct}{\mcitedefaultseppunct}\relax
\EndOfBibitem
\bibitem[Felle \latin{et~al.}(2015)Felle, Huwer, Stevenson, Skiba-Szymanska,
  Ward, Farrer, Penty, Ritchie, and Shields]{Felle.2015}
Felle,~M.; Huwer,~J.; Stevenson,~R.~M.; Skiba-Szymanska,~J.; Ward,~M.~B.;
  Farrer,~I.; Penty,~R.~V.; Ritchie,~D.~A.; Shields,~A.~J. {Interference with a
  quantum dot single-photon source and a laser at telecom wavelength}.
  \emph{{Applied Physics Letters}} \textbf{2015}, \emph{107}, 131106\relax
\mciteBstWouldAddEndPuncttrue
\mciteSetBstMidEndSepPunct{\mcitedefaultmidpunct}
{\mcitedefaultendpunct}{\mcitedefaultseppunct}\relax
\EndOfBibitem
\bibitem[Xiang \latin{et~al.}(2020)Xiang, Huwer, Skiba-Szymanska, Stevenson,
  Ellis, Farrer, Ward, Ritchie, and Shields]{Xiang.2020}
Xiang,~Z.-H.; Huwer,~J.; Skiba-Szymanska,~J.; Stevenson,~R.~M.; Ellis,~D.
  J.~P.; Farrer,~I.; Ward,~M.~B.; Ritchie,~D.~A.; Shields,~A.~J. {A tuneable
  telecom wavelength entangled light emitting diode deployed in an installed
  fibre network}. \emph{{Communications Physics}} \textbf{2020}, \emph{3},
  441\relax
\mciteBstWouldAddEndPuncttrue
\mciteSetBstMidEndSepPunct{\mcitedefaultmidpunct}
{\mcitedefaultendpunct}{\mcitedefaultseppunct}\relax
\EndOfBibitem
\bibitem[Schlehahn \latin{et~al.}(2015)Schlehahn, Kr{\"u}ger, Gschrey, Schulze,
  Rodt, Strittmatter, Heindel, and Reitzenstein]{Schlehahn.2015}
Schlehahn,~A.; Kr{\"u}ger,~L.; Gschrey,~M.; Schulze,~J.-H.; Rodt,~S.;
  Strittmatter,~A.; Heindel,~T.; Reitzenstein,~S. {Operating single quantum
  emitters with a compact Stirling cryocooler}. \emph{{The Review of scientific
  instruments}} \textbf{2015}, \emph{86}, 013113\relax
\mciteBstWouldAddEndPuncttrue
\mciteSetBstMidEndSepPunct{\mcitedefaultmidpunct}
{\mcitedefaultendpunct}{\mcitedefaultseppunct}\relax
\EndOfBibitem
\bibitem[Musia{\l} \latin{et~al.}(2020)Musia{\l}, {\.Z}o{\l}nacz, Srocka,
  Kravets, Gro{\ss}e, Olszewski, Poturaj, W{\'o}jcik, Mergo, Dybka, Dyrkacz,
  D{\l}ubek, Lauritsen, B{\"u}lter, Schneider, Zschiedrich, Burger, Rodt,
  Urba{\'n}czyk, S{\k{e}}k, and Reitzenstein]{Musia.2020}
Musia{\l},~A. \latin{et~al.}  {Plug{\&}Play Fiber--Coupled 73 kHz
  Single--Photon Source Operating in the Telecom O--Band}. \emph{{Advanced
  Quantum Technologies}} \textbf{2020}, \emph{3}, 2000018\relax
\mciteBstWouldAddEndPuncttrue
\mciteSetBstMidEndSepPunct{\mcitedefaultmidpunct}
{\mcitedefaultendpunct}{\mcitedefaultseppunct}\relax
\EndOfBibitem
\bibitem[Gao \latin{et~al.}(2022)Gao, Rickert, Urban, Gro{\ss}e, Srocka, Rodt,
  Musia{\l}, {\.Z}o{\l}nacz, Mergo, Dybka, Urba{\'n}czyk, Sek, Burger,
  Reitzenstein, and Heindel]{Gao.2022}
Gao,~T.; Rickert,~L.; Urban,~F.; Gro{\ss}e,~J.; Srocka,~N.; Rodt,~S.;
  Musia{\l},~A.; {\.Z}o{\l}nacz,~K.; Mergo,~P.; Dybka,~K.; Urba{\'n}czyk,~W.;
  Sek,~G.; Burger,~S.; Reitzenstein,~S.; Heindel,~T. {A quantum key
  distribution testbed using a plug{\&}play telecom-wavelength single-photon
  source}. \emph{{Applied Physics Reviews}} \textbf{2022}, \emph{9},
  011412\relax
\mciteBstWouldAddEndPuncttrue
\mciteSetBstMidEndSepPunct{\mcitedefaultmidpunct}
{\mcitedefaultendpunct}{\mcitedefaultseppunct}\relax
\EndOfBibitem
\bibitem[Liu \latin{et~al.}(2017)Liu, Wei, Su, Su, Ma, Chen, Ni, Niu, Yu, Wei,
  Wang, and Yu]{Liu.2017b}
Liu,~S.; Wei,~Y.; Su,~R.; Su,~R.; Ma,~B.; Chen,~Z.; Ni,~H.; Niu,~Z.; Yu,~Y.;
  Wei,~Y.; Wang,~X.; Yu,~S. {A deterministic quantum dot micropillar single
  photon source with 65{\%} extraction efficiency based on fluorescence imaging
  method}. \emph{{Scientific Reports}} \textbf{2017}, \emph{7}, 13986\relax
\mciteBstWouldAddEndPuncttrue
\mciteSetBstMidEndSepPunct{\mcitedefaultmidpunct}
{\mcitedefaultendpunct}{\mcitedefaultseppunct}\relax
\EndOfBibitem
\end{mcitethebibliography}

\end{document}